\documentclass[twocolumn,showpacs,unsortedaddress,showkeys,preprintnumbers,amsmath,amssymb]{revtex4}
\usepackage{graphicx}
\usepackage{dcolumn}
\usepackage{bm}

\begin{document}

\title{Nonlinear Statistical Modelling and Model Discovery for Cardiorespiratory Data}
\author{V.N. Smelyanskiy$^1$}
%\email{Vadim.N.Smelyanskiy@nasa.gov}
\author{D.G. Luchinsky$^{1,2}$}
\author{M.M. Millonas$^1$}
\author{A. Stefanovska$^3$}
\author{P.V.E. McClintock$^2$}

\affiliation{$^1$NASA Ames Research Center, Mail Stop 269-2,
Moffett Field, CA 94035, USA}

\affiliation{$^2$Department of Physics, Lancaster University,
Lancaster LA1 4YB, UK}

\affiliation{$^3$Faculty of Electrical Engineering, University of
Ljubljana, Tr\v{z}a\v{s}ka 25, 1000 Ljubljana, Slovenia.}

\date{\today}

\begin{abstract}
We present a Bayesian dynamical inference method for
characterizing cardiorespiratory (CR) dynamics in humans by
inverse modelling from blood pressure time-series data. This new
method
 is applicable to a broad range of stochastic dynamical models, and can be implemented
 without severe computational demands. A simple nonlinear dynamical model is found
that describes a measured blood pressure time-series in the
primary frequency band of the CR dynamics. The accuracy of the
method is investigated using surrogate data with parameters close
to the parameters inferred in the experiment. The connection of
the inferred model to a well-known beat-to-beat model of the
baroreflex is discussed.
\end{abstract}


\maketitle

\section{Introduction}
 \label{s:introduction}

Model identification is an important method used in both
fundamental and applied
research~\cite{Berger:89,Kaplan:95,Mullen:97,Mukkamala:99,Mukkamala:01,Nollo:01,Chon:96,Chon:03}
on the human cardiovascular system (CVS). Because of the
complexity of CVS dynamics and the multiplicity of its mechanism,
it is inherently difficult or impossible to isolate and study
individual response mechanisms
 in the intact
organism~\cite{Jordan:95}. In such cases mathematical models of
cardiovascular control that are consistent with the experimental
data can provide valuable insights~\cite{Seidel:95,Seidel:98a}.
Altered dynamics of the cardiovascular system is associated with a
range of cardiovascular diseases and with increased mortality, and
it is hoped that dynamical metrics will provide new means of
evaluating autonomic activity, and eventually form the basis for
diagnostic tests for many conditions (see
e.g.~\cite{Berntson:97,Malpas:02,Majercak:02}).

Despite the fact that most cardiovascular controls are
demonstrably
nonlinear~\cite{Seidel:95,Sato:99,Stefanovska:99a,Akselrod:00,Leeuwen:00,Malpas:01,Chon:03}
and are perturbed by stochastic
inputs~\cite{Stefanovska:01b,Stanley:02,Chon:03}, oversimplified
assumptions of model
linearity~\cite{Berger:89,Mullen:97,Mukkamala:99,Mukkamala:01,Nollo:01,Andrew:01,Chon:04}
and/or determinism~\cite{Seidel:95,Akselrod:00} are often made in
order to make some progress in cardiovascular system
identification. Such choices are often influenced more by the
availability of certain statistical tools and methodologies than
by biophysical or medical considerations. It is very desirable to
develop reliable methods of system identification that do not have
such limitations, and are capable of treating more realistic
models. Such models could be used to relate difficult-to-access
parameters to non-invasively-measured data~\cite{Seidel:98a}.

While a number of numerical schemes have been proposed recently to
deal with different aspects of the inverse problem using linear
approximations~\cite{Berger:89,Taylor:01,Mukkamala:01,Chon:01a},
or semi-quantitative estimations of either the strength of some of
the nonlinear terms~\cite{Jamsek:03} or the directionality of
coupling~\cite{Rosenblum:02,Palus:01},  inverse cardiovascular
inverse problems remain difficult because of the complexity and
nonlinearity of the cardiovascular interactions, as well as the
stochasticity of many dynamical inputs to the system. The problem
of nonlinear cardiovascular system identification has been
addressed in a number of
publications~\cite{Chon:96,Chon:01b,Chon:03}. Nonlinearities
generally require the use of more complex and involved numerical
techniques~\cite{McSharry:99a,Heald:00,Meyer:00a,Meyer:01,Rossi:02a,Friedrich:98,Friedrich:03a},
while the presence of dynamical noise in continuous systems can
introduced systematic errors in the estimation of the model
parameters~\cite{Rossi:02,Smelyanskiy:submitted}. Analogous
difficulties arise in a broad range of problems in many scientific
disciplines, including problems in lasers \cite{Willemsen:00a} and
molecular motors \cite{Visscher:99}, in epidemiology
\cite{Earn:00}, and in coupled matter--radiation systems in
astrophysics \cite{Christensen:02}. An obstacle to progress in
these fields is the lack of general methods of dynamical inference
for stochastic nonlinear systems. Accordingly the methods
described in this paper should be of broad interdisciplinary
interest.

In this paper we introduce a novel method for the analysis of
cardiorespiratory dynamics within a nonlinear Bayesian framework
for the inference of stochastic dynamical
systems~\cite{Smelyanskiy:submitted}. This method is applied to
the analysis of a univariate blood pressure time-series where a
simple nonlinear dynamical model based on coupled nonlinear
oscillators~\cite{Saul:89a,Stefanovska:99a,Stefanovska:01a}. is
found that describes time-series data in the relevant frequency
range. The accuracy of the method is investigated using surrogate
data with parameters close to the parameters inferred in the
experiment, and the connection of this model to a well-known
beat-to-beat model of the baroreflex is discussed.

\section{The methodolgy}
 \label{s:choice}

 In the methodological framework presented here there are
three essential steps: (i)  the input data is prepared, (ii) a
parameterized class of models chosen that describes the data, and
(iii) the parameters of this model are inferred from the
time-series data.

\subsection{Data}

\begin{figure*}%[h]
   %\begin{center}
 \includegraphics[width=17cm,height=13cm]{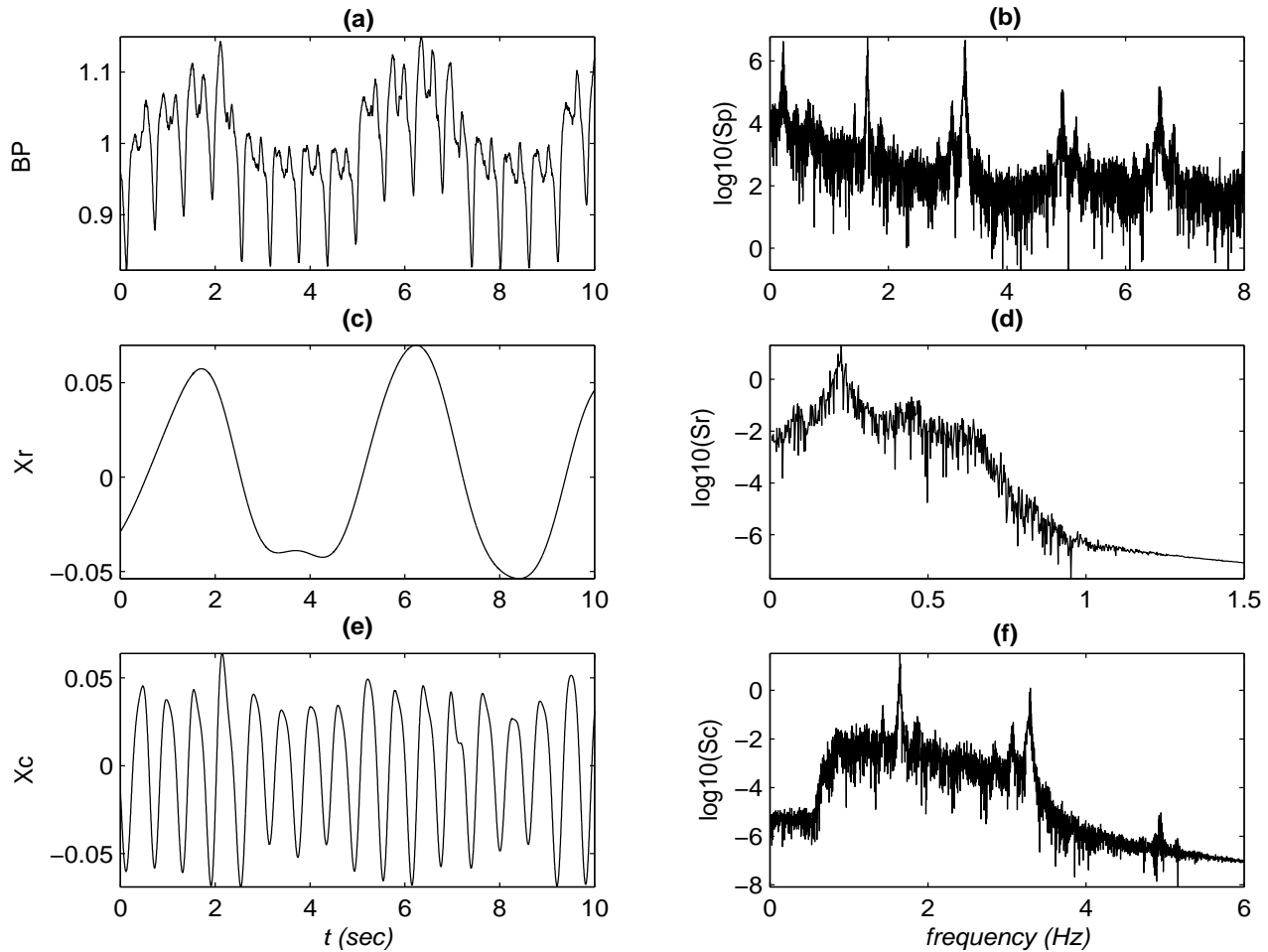}
  % \end{center}
 \caption{\label{filters}  Data from record 24 time series of the
MGH/MF Waveform Database available at www.physionet.org. (a)
Original time series of the central venous blood pressure and (b)
Power spectrum of original time series. (c) Respiratory component
produced by filtering the blood pressure time series with a 0.06
Hz, Order 2, zero-phase, high-pass Butterworth filter and a 0.6
Hz, order 12, zero phase low-pass Butterworth filter, and (d) the
power spectrum of the respiratory component. (e) Cardiac component
produced by filtering the blood pressure time series with a
0.8-3.0 Hz  Hz, order 8, zero-phase, band-pass Butterworth filter
and (d) the power spectrum of the cardiac component. The chosen
frequency range of of the components selecting according to the
discussion in the text.}
 \end{figure*}

Here we worked with a particular recording of the central venous
blood pressure, a sample of which is shown in the Fig.
\ref{filters}(a). A feature this blood pressure (BP) time-series
is the presence of the two oscillatory components at frequencies
approximately $f_r\approx 0.2$ Hz and $f_c\approx 1.7$ Hz
corresponding to the respiratory and cardiac oscillations. It can
also be clearly seen from the spectra that the nonlinear terms
including terms of nonlinear cardiorespiratory interaction
(corresponding to the side peaks) are very visible in this sample.
We note that the relative intensity and position of the cardiac
and respiratory components vary strongly from sample to sample
with average frequency of the respiration being around $0.3$ Hz
and of the heart beat being around $1.1$ Hz.

%----------------------------------------------------
%----------------------------------------------------

In preparing cardiovascular data for model identification one has
to bear in mind that the CVS power spectra reflect a large variety
of complex cardiovascular interactions seen as peaks and other
features in a very broad frequency
range~\cite{Circulation:96,Stefanovska:97a,Taylor:98,Stefanovska:01b}.
In order to make sense of these multi-scale phenomena parametric
modelling is usually restricted to a specific part of the power
spectrum. It is clear that in modelling the cardiorespiratory
interaction the frequency range of modelling must include at least
main harmonics of cardiac and respiratory oscillations $f_c$ and
$f_r$ and their combinational frequencies. Moreover, as was
pointed out already by Womersley (see e.g.~\cite{Milnor:89} cf.
also with~\cite{Javorka:02}) locally measured blood pressure
signals resembles a steady-state oscillations and the sum of the
first three harmonics  contains more then 70\% of the total signal
variance. Therefore, it is desirable that at least three harmonics
(see also discussion below) of the basic frequencies of the
respiratory and cardiac oscillations are included into the
frequency range of modelling.

 \subsection{Models}

When one considers modelling the cardiovascular system, one
usually envisions constructing a model based on biophysical
principles that is capable of generating  solutions that
reproduce, to some degree, the data: the {\em forward modelling}
problem (see
e.g.~\cite{DeBoer:87,Seidel:95,Akselrod:00,Stanley:02}). One may
also consider the {\em inverse modelling} problem, in which models
are built that describe measured data (see
e.g.~\cite{Berger:89,Mullen:97,Mukkamala:99,Mukkamala:01,Chon:01b}).
Both approaches have proven  useful in the context of the
cardiovascular research with forward approach providing valuable
insight into the system and its causal relationships, and the
inverse approach providing a useful means of intelligent patient
monitoring of cardiovascular function.

As a third alternative one may try to bridge the two approaches by
building a model that accurately reproduces the experimental
observations while at the same time is based on the biophysical
principles of circulation. In such a case the form of the
mathematical model is taken from biophysical principles, with its
component parts corresponding to a greater or lesser degree to
specific biophysical mechanisms, while the values of some or all
of the parameters of the mathematical model are inferred directly
from from the data. In such a case it is to be hoped that
information with direct biophysical significance, and not mere
mathematical or statistical characterizations can be inferred from
the data.

Many studies have been carried out to explore the physiological
mechanisms underlying cardiorespiratory
interactions~\cite{Taylor:99,Koepchen:84,Melcher:76}. The most
important ones are the modulation of cardiac filling pressure by
respiratory movements~\cite{Abel:69}, the direct respiratory
modulation of parasympathetic and sympathetic neural activity in
the brain stem ~\cite{Gilbey:84}, and the respiratory modulation
of the baroreceptor feedback control~\cite{Glass:88}. A common
 feature that these mechanisms is that they are nonlinear,
have a dynamical (or memory) component, and are subject to
exogenous
fluctuations~~\cite{Saul:89a,Braun:98,Stefanovska:99a,Suder:98,Novak:93,Chon:96,Kanters:97}.

A simple beat-to-beat model describing the cardio-respiratory
systems  DeBoer et al.~\cite{DeBoer:85,DeBoer:87}. The DeBoer
model has further been elaborated recently
in~\cite{Seidel:95,Seidel:98a,Stanley:02}. Insight into
cardio-respiratory dynamics can also be gained through inverse
modelling where the cardiac and respiratory cycles are modelled in
terms of coupled nonlinear
oscillators~\cite{Saul:89a,Stefanovska:97a,Stefanovska:99a,Stefanovska:01a,Stefanovska:01b}.
In this approach spectral and synchronization features observed in
the time-series data are interpreted physiologically, and related
to the model parameters~\cite{Stefanovska:99a}. However, the
identification of the model parameters could not be
 inferred directly from the time-series data. Instead an extensive computer simulations have
been employed to establish realistic values for the model
parameters~\cite{Stefanovska:01b}.

The simplest model that could reproduce steady-state oscillations
of the blood pressure signal at two fundamental frequencies is a
system of two coupled limit cycles on a plane. According to the
results of the Poincar$\acute{e}$-Bendixson theory of planar
dynamical systems for a system to have limit cycle in a simply
connected region the divergence of the vector field must change
sign in this region (see e.g.~\cite{Arrowsmith:82}). Therefore, we
conclude, that the simplest system that can reproduced the
discussed features of the BP signal is planar systems with limit
cycles which vector field contains polynomials of the order 3.
Accordingly, we model the time-series data as a system of two
coupled oscillators with vector fields including nonlinearities
(including nonlinearities in coupling terms) up to the 3-rd order
in the form
% \begin{widetext}
\begin{eqnarray}
\label{eq:2vdp_1}
 &&\hspace{-0.5cm}\dot x_r = a_1x_r+b_1y_r,\quad\dot y_r = \sum_{i=1}^N\alpha_i\phi_i({\bf x},{\bf y}) + \sum_{j=1}^{2}\sigma_{1j}\xi_j,\\
\label{eq:2vdp_2}
 &&\hspace{-0.5cm}\dot x_c = a_2x_c+b_2y_c,\quad\dot y_c =\sum_{i=1}^N\beta_i\phi_i({\bf x},{\bf y}) + \sum_{j=1}^{2}\sigma_{2j}\xi_j,\\
 &&\langle\xi_i(t)\rangle = 0, \qquad \langle\xi_i(t)\xi_j(t')\rangle = \delta_{ij}\delta(t-t').\nonumber
\end{eqnarray}
% \end{widetext}
Here noise matrix $\sigma$ mixes zero-mean white Gaussian noises
$\xi_j(t)$, which is related to the diffusion matrix
$D=\sigma\sigma^T$. And base functions are chosen in the form
\begin{eqnarray}
\label{eq:basis}
  &&\hspace{-0.4cm}\phi=\{1, x_r, x_c, y_r, y_c, x_r^2,
      x_c^2, y_r^2, y_c^2, x_ry_r, x_cy_c, x_r^3, \nonumber\\
 &&\hspace{-0.4cm} x_c^3, x_r^2y_r, x_c^2y_c, x_ry_r^2, x_cy_c^2, y_r^3,
 y_c^3, x_rx_c, x_r^2x_c, x_rx_c^2 \}.
\end{eqnarray}
%----------------------------------------------------
 \begin{figure}[h]
   \begin{center}
 \includegraphics[width=7cm,height=5.cm]{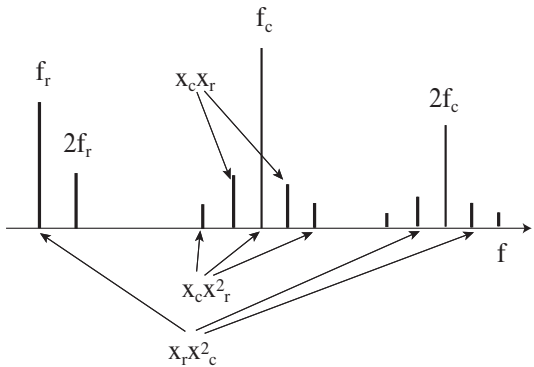}
   \end{center}
 \caption{\label{fig:summary}
Summary of the main harmonics of the cardiac and respiratory
components observed in the BP signal. The correspondence between
the nonlinear terms of the model (\ref{eq:2vdp_1}),
(\ref{eq:2vdp_2}) and the frequencies observed in the time-series
data are shown by arrows.}
 \end{figure}
%----------------------------------------------------
The restrictions imposed on the rhs of the equations for $\dot
x_r$ and $\dot x_c$ in (\ref{eq:2vdp_1}) and (\ref{eq:2vdp_2}) are
determined mainly by the fact that we have to infer four hidden
dynamical variables using univariate time-series data (see next
section for further details). The parametric representation
(\ref{eq:2vdp_1}) and (\ref{eq:2vdp_2}) covers a wide range of
models with limit cycles in the plane. In particular, with a
special choice of the model parameters it describes van der Pol or
FitzHugh-Nagumo oscillator systems that are very popular in the
context of the cardiovascular modelling. Furthermore, the choice
of the parametric model in the form (\ref{eq:2vdp_1}) and
(\ref{eq:2vdp_2}) allows one to relate them to physiological
parameters characterizing autonomous nervous system (see section
\ref{s:discussion} for the discussion). See \cite{Stefanovska:99a}
for the alternative choice and corresponding physiological
reasoning.

\subsection{Parameters}

Following the logic of the inverse modelling approach, we must
then identify the parameters ${\cal M}=\{{\bf a},{\bf
b},{\boldsymbol \alpha},{\boldsymbol \beta},D\}$ of the model
(\ref{eq:2vdp_1}), (\ref{eq:2vdp_2}) that reproduce the dynamical
and spectral features of the BP signal shown in the Fig. 1. Terms
representing nonlinear cardio-respiratory interactions are
described by the last three base functions in (\ref{eq:basis}).
The correspondence of these terms to the experimentally observed
combinational frequencies in the BP signal is summarized in the
Fig. \ref{fig:summary}. It can be seen from the figure that the
same combinational frequencies correspond to the nonlinear
coupling terms in both limit cycle systems in the model therefore
a {\em nonlinear} time-series analysis is a requirement for the
identification of such a model.

Here we show how the task of identifying a model of the
cardiovascular system based on coupled nonlinear oscillators can
be performed systematically within a Bayesian statistical
framework. In this approach parameters of the model can be
inferred directly from the time-series data. We also discuss
briefly how the links between the coupled oscillator model and
beat-to-beat models can in principle be established.

\section{Bayesian inference of stochastic nonlinear dynamical models}
 \label{s:bayesian}

Details of our new Bayesian technique can be found
elsewhere~\cite{Smelyanskiy:submitted}. Here we give a brief
description of the main steps of the algorithm.

Stochastic nonlinear dynamical models of the type
(\ref{eq:2vdp_1}), (\ref{eq:2vdp_2}) can be expressed as a
multi-dimensional nonlinear Langevin equation
\begin{equation}
    \dot{\bf x}(t) = {\bf f}({\bf x}) + {\bf \varepsilon}(t)
    = {\bf f}({\bf x}) + {\bf\sigma}{\bf \xi}(t),
    \label{eq:dynamics}
\end{equation}
\noindent where ${\bf \varepsilon}(t)$ is an additive stationary
white, Gaussian vector noise process characterized by
\begin{equation}
    \langle {\bf \xi}(t) \rangle = 0, \quad \langle {\bf \xi}(t) \, {\bf
    \xi}^{T}(t') \rangle = {\hat {\bf D}} \, \delta(t - t'),
    \label{eq:noise}
\end{equation}
where ${\hat {\bf D}}$ is a diffusion matrix.

It is assumed that the trajectory $x(t)$ of this system is
observed at sequential time instants  $\lbrace t_k; k = 0, 1,
\ldots, K \rbrace$ and a series ${\cal S} = \lbrace s_k \equiv
s(t_{k}) \rbrace$ thus obtained is related to the (unknown)
``true'' system states ${\cal X} = \lbrace x_k \equiv x( t_{k})
\rbrace$ through some conditional PDF $p_{\rm o}\left({\cal
S}|{\cal X} \right)$.

An {\it a priori} expert knowledge about the model parameters is
summarized in so-called {\em prior} PDF $p_{\textrm{pr}}({\cal
M})$. In our case we chose {\em prior} in the form of zero-mean
Gaussian distribution for model parameters and uniform
distributions for the coefficients of diffusion matrix.

If experimental time-series data ${\cal S}$ are available they can
be used to improve the estimation of the model parameters. The
improved knowledge of the models parameters is summarized in the
{\em posterior} conditional PDF $p_{\textrm{post}}({\cal M}|{\cal
S})$, which is related to {\em prior} via Bayes' theorem:
\begin{equation}
    \label{eq:Bayes}
  p_{\textrm{post}}({\cal M}|{\cal S}) = \frac{{\ell}({\cal S}|{\cal M}) \,
  p_{\textrm{pr}}({\cal M})}{\int \ell({\cal S}|{\cal M}) \, p_{\textrm{pr}}({\cal M}) \, {\rm d}{\cal M}}.
\end{equation}
Here $\ell({\cal S}|{\cal M})$, usually termed the
\emph{likelihood}, is the conditional PDF that relates
measurements ${\cal S}$ to the dynamical model. The determinant on
the right hand side of the equation (\ref{eq:Bayes}) is merely a
normalization factor. In practice, (\ref{eq:Bayes}) can be applied
iteratively using a sequence of data blocks ${\cal S},{\cal
S}^{\prime}$, etc. The posterior computed from  block ${\cal S}$
serves as the prior for the next block ${\cal S}^{\prime}$, etc.
For a sufficiently large number of observations,
$p_{\textrm{post}}({\cal M}|{\cal S},{\cal S}^{\prime},\ldots)$ is
sharply peaked at a certain most probable model $\cal M={\cal
M}^{\ast}$.

The main efforts in the research on stochastic nonlinear dynamical
inference are focused on constructing the likelihood function,
which compensates noise induced errors, and on introducing
efficient algorithms of optimization of the likelihood function
and integration of the normalization factor
(cf.~\cite{McSharry:99a,Meyer:00,Meyer:01,Rossi:02}).

In our earlier work~\cite{Smelyanskiy:submitted} a novel technique
of nonlinear dynamical inference of stochastic systems was
introduced that solves both problems. To avoid extensive numerical
methods of optimization of the likelihood function and integration
of the normalization factor we suggested to parameterize the
vector field of (\ref{eq:dynamics}) in the form
\begin{equation}
    \label{eq:model}
    {\bf f}({\bf x}) = {\hat {\bf U}}({\bf x}) \, {\bf c} \equiv {\bf f}({\bf x}; {\bf c}),
\end{equation}
\noindent where ${\hat {\bf U}}({\bf x})$ is an $N \times M$
matrix of suitably chosen basis functions $\lbrace U_{n m}({\bf
x}); \,n = 1:N,\, m = 1:M \rbrace$, and ${\bf c}$ is an
$M$-dimensional coefficient vector. An important feature of
(\ref{eq:model}) is that, while possibly highly nonlinear in ${\bf
x}$, ${\bf f}({\bf x}; {\bf c})$ is strictly linear in ${\bf c}$.

The computation of the likelihood function can be cast in the form
of a path integral over the random trajectories of the
system~\cite{Graham:77,Dykman:90}. Using the uniform sampling
scheme introduced above we can write the logarithm of the
likelihood function in the following form for sufficiently small
time step $h$ (cf.~\cite{Graham:77,Dykman:90}):
\begin{eqnarray}
 && \hspace{-0.3in}-\frac{2}{K}\log  \ell({\bf y}|{\cal M})  =
\ln\det{\hat{\bf D}}
+\frac{h}{K}\sum_{k=0}^{K-1}\left[\,{\bf v}({\bf y}_k){\bf c}\right.\label{eq:likelihood} \\
  &&\hspace{-0.2in}\left. +(\dot{\bf y}_{k}  - {\hat {\bf U}}_k \, {\bf c})^T \, {\hat {\bf D}}^{-1} \,
    (\dot{\bf y}_{k}  - {\hat {\bf U}}_k \, {\bf c}))\right]+ N\ln(2\pi h),\nonumber
\end{eqnarray}
\noindent which relates the dynamical variables ${\bf x}(t)$ of
the system (\ref{eq:dynamics}) to the observations ${\bf s}(t)$.
Here we introduce the following notations $ { \bf \hat U}_{k}
\equiv {\bf \hat U}({\bf y}_{k})$, $\dot {\bf y}_{k}\equiv h^{-1}
({\bf y}_{k+1}-{\bf y}_k)$ and vector ${\bf v}({\bf x})$ with
components
\[\textrm{v}_{m}({\bf x})=\sum_{n=1}^{N}\frac{\partial U_{n\,m}({\bf
x})}{\partial x_n},\quad m=1:M.\] The vector elements $\lbrace c_m
\rbrace$ and the matrix elements $\lbrace D_{n n'} \rbrace$
together constitute a set ${\cal M} = \lbrace {\bf c}, {\hat {\bf
D}} \rbrace$ of unknown parameters to be inferred from the
measurements ${\cal S}$.

Choosing the prior PDF in the form of Gaussian distribution
\begin{equation}
    p_{\textrm{pr}}({\cal M}) = \sqrt{\frac{\det({\bf \hat  \Sigma}^{-1}_{\textrm{pr}})}{(2\pi)^M}}
    \exp\left(-\frac{1}{2} ({\bf c}-{\bf c}_{\textrm{pr}})^{T}{\bf \hat
    \Sigma}^{-1}_{\textrm{pr}}({\bf c}-{\bf c}_{\textrm{pr}})\right).
    \label{eq:prior}
\end{equation}
and substituting $p_{\textrm{pr}}({\cal M})$ and the  likelihood
$\ell({\cal S}|{\cal M})$ into (\ref{eq:Bayes}) yields the
posterior $p_{\textrm{post}}({\cal M}|{\cal S}) ={\rm const}\times
\exp[-L({\cal M}|{\cal S})]$, where
\begin{equation}
    L({\cal M}|{\cal S})\equiv L_{\textsf{s}}({\bf c},{\bf \hat D}) =
    \frac{1}{2}\rho_{\textsf{s}}({\bf \hat  D}) - {\bf c}^{T} {\bf w}_{\textsf{s}}({\bf \hat  D}) +
     \frac{1}{2}
    {\bf c}^{T} {\bf \hat  \Xi}_{\textsf{s}}({\bf \hat  D}) {\bf c}.
    \label{eq:action}
\end{equation}
\noindent Here,  use was made of the definitions
\begin{eqnarray}
  &&\hspace{-0.25in}\rho_\textsf{s}({\bf \hat{D}}) =  h \, \sum_{k = 0}^{K - 1} \dot{
  {\bf s}}_{k}^{T}   \, {\bf \hat D}^{-1} \, \dot{{\bf s}}_{k}  + K \, \ln (\det { \bf \hat D}),
  \label{eq:defs_1} \\
 &&\hspace{-0.25in}{\bf w}_\textsf{s}({\bf \hat D}) = {\bf \hat \Sigma}_{\textrm{pr}}^{-1} \,
   {\bf c}_{\textrm{pr}} + h\sum_{k = 0}^{K - 1}\left[ {\bf \hat U}_{k}^T \,
   {\bf \hat D}^{-1} \, \dot{{\bf s}}_{k}-  \frac{{\bf v}({\bf s}_k)}{2}\right],
   \label{eq:defs_2} \\
 &&\hspace{-0.25in}{\bf \hat  \Xi}_\textsf{s}({\bf \hat  D})
    =  {\bf \hat  \Sigma}^{-1}_{\textrm{pr}} + h \, \sum_{k = 0}^{K - 1}
    {\bf \hat U}_{k}^{T} \, {\bf \hat D}^{-1} \, {\bf \hat U}_{k}.
    \label{eq:defs_3}
\end{eqnarray}

The mean values of ${\bf c}$ and ${\bf \hat D}$ in the posterior
distribution give the best estimates for the model parameters for
a given block of data ${\cal S}$ of  length $K$ and provide a
global minimum to $L_\textsf{s}({\bf c}, {\hat {\bf D}})$. We
handle this optimization problem in the following way. Assume for
the moment that ${\bf c}$ is known in (\ref{eq:action}). Then the
posterior distribution over ${\bf \hat D}$ has a mean  ${\bf\hat
D}^{\bf \prime}_{\textrm{post}}={\bf \hat
\Theta}_{\textsf{s}}({\bf c})$ that provides a  minimum to
$S_\textsf{s}({\bf c},{\bf \hat D})$ with respect to ${\bf \hat
D}={\bf \hat D}^T$.   Its matrix elements are
\begin{equation}
 \label{eq:updateD}
  \hspace{-0.01in}{\bf \hat \Theta}_{\textsf{s}}^{n n'}({\bf c})
  \equiv \frac{1}{K} \, \sum_{k=0}^{K-1} \left[ {\dot {\bf s}}_{k} -
    {\hat {\bf U}}({\bf s}_{k}) \, {\bf c} \right]_n \left[ {\dot {\bf s}}_{k} - {\hat {\bf U}}({\bf
    y}_{k}) \, {\bf c} \right]^{T}_{n'}.
\end{equation}
\noindent Alternatively, assume next that
 ${\hat {\bf D}}$ is known, and note
from (\ref{eq:action}) that in this case the posterior
distribution over ${\bf c}$ is Gaussian. Its covariance is given
by ${\bf \hat\Xi}_\textsf{s}({\bf \hat D})$ and the mean ${\bf
c}^{\prime}_{\textrm{post}}$ minimizes  $L_\textsf{s}({\bf c},{\bf
\hat D})$ with respect to ${\bf c}$
\begin{equation}
%{\bf \mu}_\textsf{s}({\bf \hat D})
{\bf c}^{\prime}_{\textrm{post}}={\hat {\bf
\Xi}}^{-1}_\textsf{s}({\bf \hat D}){\bf w}_\textsf{s}({\bf \hat
D}).\label{eq:updateC}
\end{equation}
\noindent We repeat  this two-step optimization procedure
iteratively, starting from some prior values ${\bf
c}_{\textrm{pr}}$ and ${\bf \hat \Sigma}_{\textrm{pr}}$.

\section{Estimation of parameters of cardiorespiratory interaction from univariate time-series data (real data)}
 \label{s:cvs}

In order to apply algorithm (\ref{eq:defs_2})-(\ref{eq:updateC})
for the identification of the model of nonlinear
cardio-respiratory dynamics (\ref{eq:2vdp_1}), (\ref{eq:2vdp_2})
from the univariate BP time-series of the type shown in Fig. 1(a)
we have to extract time-series data corresponding to the four
dynamical variable in the model.  Accordingly we divide the total
spectrum into a low-frequency respiratory component $s_r(t)$ and
high-frequency cardiac component $s_c(t)$ as is shown in Fig. 1(c)
and (e).

A discussion of the physiological relevance of this spectral
separation can be found
in~\cite{Stefanovska:99a,Stefanovska:01a}).  However, it is
perfectly correct to consider this separation a mathematical {\it
ansatz}.   Physiological considerations need only come into play
at the point where we attempt place a specific biophysical
interpretation of the model elements. The parameters of the
filters (see Fig. 1 caption) were chosento preserve the 2-nd and
3-rd harmonics of these signals. Then two $x_r(t)$ and $x_c(t)$
dynamical variables of the model (\ref{eq:2vdp_1}),
(\ref{eq:2vdp_2}) can be identified with introduced above
two-dimensional time-series of observations ${\bf s}(t)=\{s_r(t),
s_c(t)\}$. The remaining two dynamical variables ${\bf
y}(t)=\{y_r(t),y_c(t)\}$ can be related to the observations
$\{{\bf s}(t_k)\}$  as follows
\begin{eqnarray}
 \label{eq:embedding}
 b_ny_n(t_k) = \frac{s_n(t_k+h)-s_n(t_k)}{h} +a_{n}s_n(t_k),
\end{eqnarray}
where $n=r,c$. The relation (\ref{eq:embedding}) is a special form
of embedding that allows one to infer a wider class of dynamical
models of the cardiorespiratory interactions including models in
the form of FitzHugh-Nagumo oscillators. It is clear now that we
have introduced the restrictions on the form of the rhs of the
first equations in (\ref{eq:2vdp_1}), (\ref{eq:2vdp_2}) to reduce
the number of parameters of embedding that have to be selected to
minimize the cost (\ref{eq:action}) and provide the best fit to
the measured time series $\{ {\bf s}(t_k)\}$. The corresponding
simplified model of the nonlinear interaction between the cardiac
and respiratory limit cycles can now be written in the form
corresponding to parametrization (\ref{eq:model}) as follows
\begin{eqnarray}
 &&\dot {\bf y} = {\bf \hat U}({\bf s},{\bf y}){\bf c} + {\boldsymbol \xi}(t),
 \label{eq:cardiorespiratory}
\end{eqnarray}
\noindent where $ {\boldsymbol \xi}(t)$ is a two-dimensional
Gaussian white noise with correlation matrix ${\bf\hat D}$, and
the matrix ${\hat {\bf U}}$ will have the following block
structure
%\begin{widetext}
\begin{eqnarray}
 \label{eq:block_structure_1}
&&\hspace{-0.25in}{\hat {\bf U}}=\left[
  \left[\begin{array}{ll}
    \phi_1&0 \\
    0&\phi_1 \\
  \end{array}\right]   \ldots
  \left[\begin{array}{ll}
    \phi_2&0 \\
    0&\phi_2 \\
  \end{array}\right] \ldots
  \left[\begin{array}{ll}
    \phi_B&0 \\
    0&\phi_B \\
  \end{array}\right]
  \right].
\end{eqnarray}
%\end{widetext}
Here $B=22$ diagonal blocks of size $2\times 2$ formed by the
basis functions given in (\ref{eq:basis}) and the vector of
unknown parameters ${\bf c}$ has the length $M=2B$.

\begin{figure}
\includegraphics[width=8.cm,height=12cm]{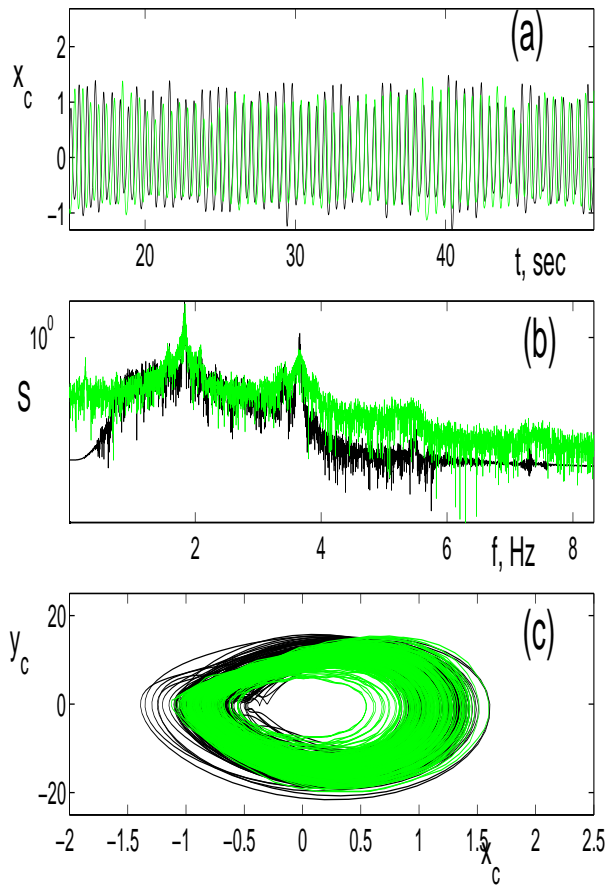}
\caption{\label{fig:cardiac} (a) Time series of the cardiac
oscillations $x_c(t_n)=s_c(t_n)$ in arbitrary units (black line)
obtained from central venous blood pressure. The sampling rate was
90 Hz after resampling of the original signal. Inferred time
series of the cardiac oscillator (green line). (b) Power spectrum
of the cardiac oscillations obtained from the real data (black
line). Power spectrum of the inferred oscillations (green line).
(c) Limit cycle of the cardiac oscillations $(x_c(n),y_c(n)$
obtained from real data as described in the text (black line).
Limit cycle of inferred oscillations (green line).}
\end{figure}

\begin{figure}
\includegraphics[width=8.cm,height=12cm]{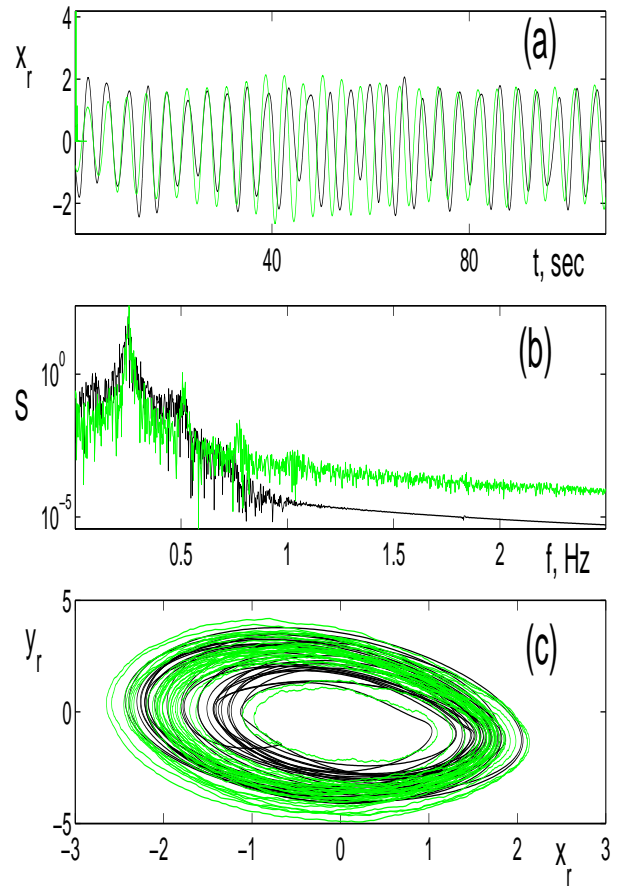}
\caption{\label{fig:respiration} (a) Time series of the
respiratory oscillations $x_r(t_n)=s_r(t_n)$ in arbitrary units
(black line) obtained from central venous blood pressure. The
sampling rate was 90 Hz after resampling of the original signal.
Inferred time series of the respiratory oscillator (green line).
(b) Power spectrum of the respiratory oscillations obtained from
the real data (black line). Power spectrum of the inferred
oscillations (green line). (c) Limit cycle of the respiratory
oscillations $(x_c(n),y_c(n)$ obtained from real data as described
in the text (black line). Limit cycle of inferred oscillations
(green line).}
\end{figure}

Finally, the model (\ref{eq:cardiorespiratory}),
(\ref{eq:block_structure_1}) has to be inferred using method
described in the previous section. The comparison between the time
series of the inferred and actual cardiac oscillations is shown in
Fig. \ref{fig:cardiac}.
%---------------------------------------
%------------------------------------------
Similar results are obtained for the respiratory oscillator as
shown in the Fig. \ref{fig:respiration}.
%---------------------------------------
%------------------------------------------
In particular, the parameters of the nonlinear coupling and of the
noise intensity  of the cardiac oscillations have been estimated
to have the following values $\beta_{20}=2.2, \beta_{21}=0.27$,
$\beta_{22}=-8.67$, and $\langle\xi_c^2(t)\rangle=8.13$. While
parameters of the coupling of respiratory oscillations to the
cardiac oscillations were estimated to have the following values
$\alpha_{20}=0.12, \alpha_{21}=0.048$, $\alpha_{22}=-0.066$, and
$D_{11}=0.18$ . Consistent with expectations, in all experiments
the  parameters of the nonlinear coupling are more then one order
of magnitude higher for the cardiac oscillations as compared to
their values for the respiratory oscillations reflecting the fact
that respiration strongly modulates cardiac oscillations, while
the opposite effect of the cardiac oscillations on  respiration is
weak.

We have shown that it possible using these methods to
simultaneously infer the coupling strengths and noise parameters
nonlinear cardio-respiratory dynamics directly from a
non-invasively measured time series. We view this demonstration of
principle as a first step towards the practical use this technique
for cardio-respiratory modelling and and in clinical applications.
A number of very important physiological and mathematical issues
arise in relation to the application of this new technique to
specific problems.  We hope to address many of these in future
publications. In what follow here we specifically consider
 the problem of estimating of the accuracy of the method,
and begin the discussion of the connection between inferred
parameters and the indexes of autonomous cardiovascular controls.

\section{Validation of the method using surrogate time-series data}
 \label{s:surrogate}

%---------------------------------------
\begin{figure}
\includegraphics[width=8.cm,height=6cm]{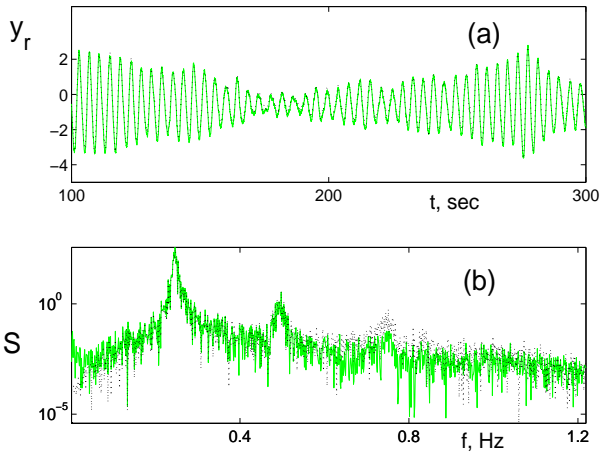}
\caption{\label{fig:surr_embedding} (a) The velocity of the
respiratory component of oscillations of the original signal
$y_r(t)$ (green line) is compared with the signal $\tilde
y_r(t_k)$ (black dashed line) obtained as a result of filtration
of $s(t)$ followed by the embedding $b_1\tilde y_r(t_k) =
(s_r(t_k+h)-s_r(t_k))/h +a_2s_r(t_k)$. (b) Power spectra of the
original velocity of the respiratory component $y_r(t_k)$ (green
line) is shown in comparison with the power spectrum of the
recovered signal $\tilde y_r(t_k)$ (black dashed line).}
\end{figure}
%------------------------------------------

It is desirable to check performance of the method on surrogate
time-series data obtained by numerically simulating the model
(\ref{eq:2vdp_1}), (\ref{eq:2vdp_2})with the parameters measured
with the CVS data.

To this end we consider a surrogate signal $x(t)=x_r(t)+x_c(t)$ as
a time-series data input $s(t)$ for the inference. Here $x_r(t)$,
$x_c(t)$ are obtained using numerical simulations of the model
(\ref{eq:2vdp_1}), (\ref{eq:2vdp_2}) with the parameters taken
from the inference of the experimental BP signal described in the
previous section.

First we verify that the decomposition of the input signal $s(t)$
into low-frequency  $\tilde s_r$ and high-frequency $\tilde s_c$
harmonics using two band-pass Butterworth filters and subsequent
application of the embedding procedure (\ref{eq:embedding}) allows
one to reconstruct the original signal. In the Fig.
\ref{fig:surr_embedding} we compare the velocity of the
respiratory component of the original signal $y_r(t)$ with the the
reconstructed velocity $\tilde y_r(t)$. Similar results are
obtained for the reconstruction of the high-frequency component.
We notice in particular that the noise introduced by embedding can
be neglected since it is more then order of magnitude smaller then
the dynamical noise in the signal.
%---------------------------------------
\begin{figure}
\includegraphics[width=8.cm,height=12cm]{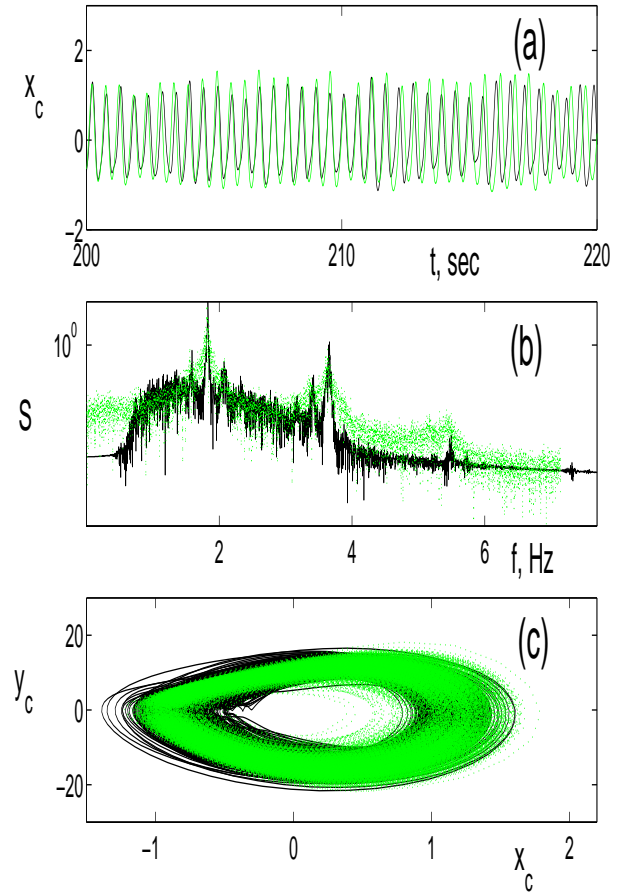}
\caption{\label{fig:surr_total} (a) Surrogate time series of the
respiratory oscillations $x_r(t_n)$ in arbitrary units (black
line) obtained from model (\ref{eq:2vdp_1}), (\ref{eq:2vdp_2}).
Inferred time series of the cardiac oscillator (green line). (b)
Power spectrum of the surrogate respiratory oscillations (black
line). Power spectrum of the inferred oscillations (green dashed
line). (c) Limit cycle of the surrogate respiratory oscillations
$(x_c(n),y_c(n)$ (black line). Limit cycle of inferred
oscillations (green dashed line).}
\end{figure}
%------------------------------------------

Now we can apply inference procedure described in the previous
section to estimate nonlinear coupling parameters of the model
from the univariate surrogate time-series data. The results of the
estimation are summarized in the Table \ref{tab:1}.
\begin{table}[ht]
   \begin{tabular}{|c|c|c|c|c|c|c|c|}\hline
   $\alpha_{20}$ & $\beta_{20}$ & $\alpha_{21}$ & $\beta_{21}$ & $\alpha_{22}$ & $\beta_{22}$ & $D_{11}$  & $D_{22}$\\
        \hline
   0.12  & 2.2  &  0.048  &  0.27 &  -0.066 &  -8.67  &  0.18  &  8.13\\
        \hline
    0.18  &  6.32  &  0.011  &  0.49  &  0.053  &  6.03  &  0.017  &  3.44\\
        \hline
   51.2\%  &  186.8\%  &   75.9\%  &  102.7\%  &  27.9\% & 30.6\% &  90.8\% &  57.7\%\\
        \hline
        \end{tabular}
    \caption{Absolute values of the coefficients of nonlinear cardiorespiratory interactions
    corresponding to the last three base functions (\ref{eq:basis})
    $\{x_rx_c,x_r^2x_c,x_rx_c^2\}$. Coefficients $\{\alpha_i\}$ correspond to the respiration
    coupling to cardiac  rhythm. Coefficients $\{\beta_i\}$ correspond to the
    cardiac oscillation coupling to respiration. For each set of coefficients the actual values
    (top row) are compared with the mean inferred values obtained from 100 blocks of time-series
    data $x(t)=x_r(t)+x_c(t)$ with 50000 points in each block and sampling time 0.02 sec
    (middle row). The error of estimation is shown in the bottom line.}
    \label{tab:1}
\end{table}
%    0.1192    2.2044    0.0475    0.2708   -0.0664   -8.6693    0.1794    8.1268
%    0.1802    6.3224    0.0114    0.4857    0.0526    6.0267    0.0166    3.4350
%   51.1581  186.8061   75.9206  102.6755  -27.8744  -30.5633   90.7533   57.7321
It can be seen from the table that the method allows one to
estimate correct order of the absolute values of the nonlinear
coupling parameter. For some parameter the accuracy of the
estimation is much better, but in practice the correct values are
not know. Therefore we conclude that the accuracy of the
estimation of the absolute values of parameters of coupling of two
limit cycle systems from univariate time-series data is within the
order of magnitude.

\begin{table}[ht]
   \begin{tabular}{|c|c|c|c|c|c|c|c|}\hline
   $\alpha_{20}$ & $\beta_{20}$ & $\alpha_{21}$ & $\beta_{21}$ & $\alpha_{22}$ & $\beta_{22}$ & $D_{11}$  & $D_{22}$\\
        \hline
   0.12  &  2.20  &  0.048  &  0.27 &  -0.066 &  -8.67  &  0.18  &  8.13\\
        \hline
   0.12  &  2.41  &  0.048  &  0.28 &  -0.070 &  -8.61  &  0.18  &  8.14\\
        \hline
    2.9\%  &  9.3\%  &  1.8\%  &  5.6\% &  5.2\% &  0.7\%  &   0.2\%  &  0.2\% \\
        \hline
        \end{tabular}
    \caption{Absolute values of the coefficients of nonlinear cardiorespiratory interactions
    corresponding to the last three base functions (\ref{eq:basis})
    $\{x_rx_c,x_r^2x_c,x_rx_c^2\}$. Coefficients $\{\alpha_i\}$ correspond to the respiration
    coupling to cardiac  rhythm. Coefficients $\{\beta_i\}$ correspond to the
    cardiac oscillation coupling to respiration. For each set of coefficients the actual values
    (top row) are compared with the mean inferred values obtained from 100 blocks of time-series
    data $\{x_r(t), x_c(t), y_r(t), y_c(t)\}$ with 160000 points in each block and sampling
    time 0.01 sec (middle row). The error of estimation is shown in the bottom line.}
    \label{tab:2}
\end{table}
%    0.1192    2.2044    0.0475    0.2708   -0.0664   -8.6693    0.1794    8.1268
%    0.1210    2.4094    0.0482    0.2774   -0.0698   -8.6128    0.1797    8.1396
%    2.8883    9.3006    1.8248    5.5731   -5.2345   -0.7057    0.1760    0.1582
Similar results are obtained for the estimation of other
parameters of the model. Using values of the model parameters
estimated from the univariate surrogate data one can reconstruct
very closely the dynamical and spectral features of the original
system as shown in the Fig. \ref{fig:surr_total}. The largest
errors of estimations are obtained for the values of the noise
intensity as shown in two last columns of the Table
\ref{fig:surr_total}. This result can be easily understood taking
into account that filtration of the signals has the strongest
effect on the noise spectrum of the system. However, the
filter-induced errors are systematic and can be corrected using
test with surrogate data.

The main source of error is related to the spectral decomposition
of the univariate data and it is therefore systematic. To
illustrate this point we use the original surrogate time-series
data  $\{x_r(t), x_c(t), y_r(t), y_c(t)\}$ for two coupled
oscillators to infer parameters of the model (\ref{eq:2vdp_1}),
(\ref{eq:2vdp_2}). The results of inference of the coupling
parameters are shown in the Tab. \ref{tab:2}. It can be seen that
the values of the parameters can be estimated with relative error
better then 10\%. In particular, the relative error of estimation
of the noise intensity is now better then 4\%. The accuracy of the
estimation can be further improved by increasing the total time of
observation of the system dynamics as explained
in~\cite{Smelyanskiy:submitted}.

These results should be compared with semi-quantitative
estimations of either relative strength of some of the nonlinear
terms~\cite{Jamsek:03} or the directionality of
coupling~\cite{Rosenblum:02,Palus:01} from bivariate time-series
data. It becomes clear that our algorithm provides an alternative
effective approach to the solution of the problem of analysis of
cardiovascular coupling.  In particular, the results of this
section validate the application of the method to the experimental
time-series cardiovascular data and demonstrate that the it is
indeed possible to estimate simultaneously the strength,
directionality and the noise of nonlinear cardiorespiratory
coupling form the univariate blood pressure signal. The accuracy
of the estimation is within the order of the magnitude. The main
source of errors is the decomposition of the univariate signal
into two oscillatory components and it is, therefore, systematic.
Using this fact one can introduce systematic corrections to
improve the results of the estimation of the parameters of the CR
interaction from the experimentally measured BP signal.

\section{Discussion}
 \label{s:discussion}

It is important to establish a relationship between the model
parameters and physiological parameters of the cardiovascular
system. A beat-to-beat model describing the relationships between
blood pressures and respiration in simple, but physiologically
meaningful terms is the DeBoer model~\cite{DeBoer:85,DeBoer:87}.
While the DeBoer model cannot describe the dynamics within one
heartbeat it does incorporate several well-known physiological
laws of the cardio-respiratory system. More recent extensions and
modifications of the DeBoer model have
appeared~\cite{Seidel:95,Seidel:98a,Stanley:02}. The problem of
inverse modelling was not addressed in this earlier work. It is
therefore very desirable to connect the approach presented here
with such beat-to-beat models.

\begin{figure}
\includegraphics[width=8.cm,height=6cm]{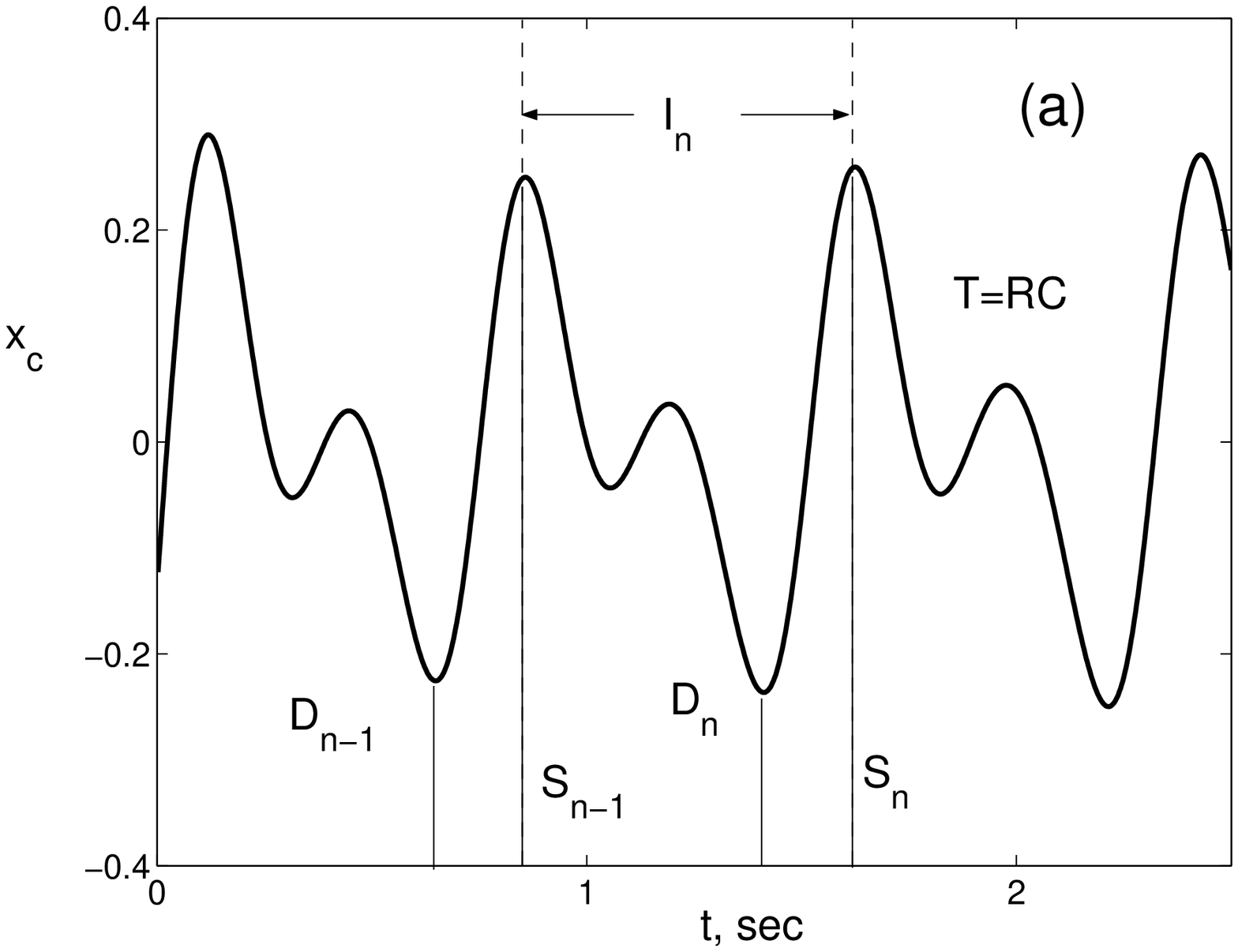}
\includegraphics[width=8.cm,height=6cm]{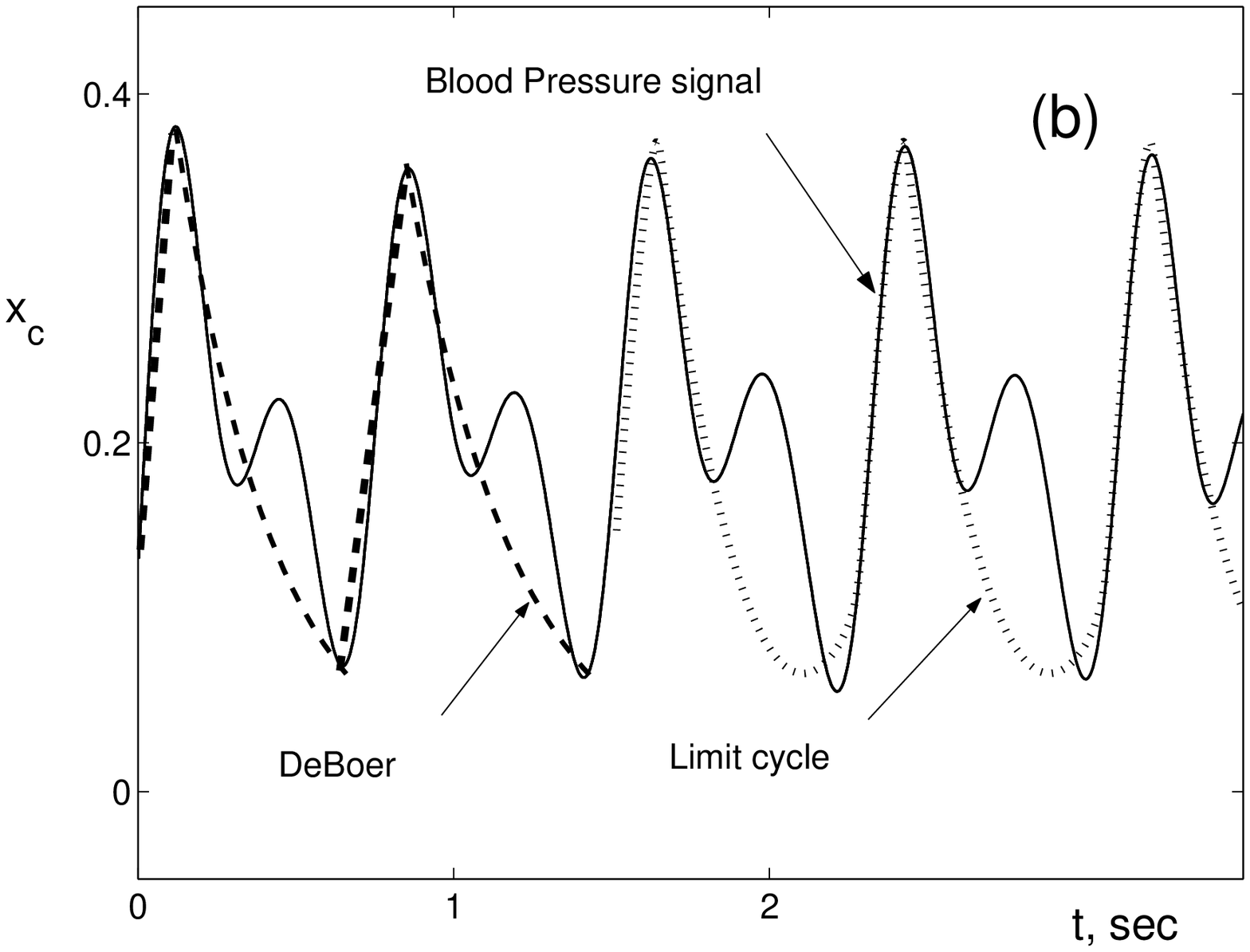}
\caption{\label{fig:DeBoer_1} (a) The BP signal in the frequency
range of cardiac oscillations (black line). Systolic pressure
($S_n$), diastolic pressure ($D_n$), RR intervals ($I_n$), and
arterial decay time ($T = RC =$) are shown for the $n$-th
heartbeat. (b) Comparison of the BP signal (thing black line) with
the approximation adopted in the DeBoer model (dashed line) and
the approximation by the FitzHugh-Nagumo model (dotted line).
Vertical scale has arbitrary units.}
\end{figure}

The DeBoer model describes the beat-to-beat evolution of the
 state variables shown in the Fig. \ref{fig:DeBoer_1}
(a): systolic pressure ($S$), diastolic pressure ($D$), RR
intervals ($I$), and arterial decay time ($T = R \times C =$
peripheral resistance $\times$ arterial compliance). Following a
brief account of the DeBoer model given in~\cite{Akselrod:00} and
neglecting for the sake of simplicity the variation of the
peripheral resistance we can write a set of corresponding
difference equations in the form
\begin{eqnarray}
 \label{eq:DeBoer_1}
  D_i &=& S_{i-1}\exp[(-2/3)I_{i-1}/T], \\
 \label{eq:DeBoer_2}
  S_i &=& D_i+\gamma I_{i-1}+C_1+A\sin(2\pi f t), \\
 \label{eq:DeBoer_3}
  I_i &=& G_vS'_{i-\tau_v} + G_{\beta}F(S',\tau_{\beta}) + C_2,
% \label{eq:DeBoer_4}
%  T_i &=& C_3 - G_{\alpha}F(S',\tau_{\alpha}).
\end{eqnarray}
Here $C_1$, $C_2$, and $C_3$ are constants and the sigmoidal
nature of the baroreceptor sensitivity is accounted for by
defining an effective Systolic pressure ($S'$)~\cite{DeBoer:87}
\begin{eqnarray}
 \label{eq:baroreflex}
S'_i = S_0 + 18 \arctan[(S-S_0)/18].
\end{eqnarray}
The first equation (\ref{eq:DeBoer_1}) follows from the Windkessel
model of the circulation, while the second equation
(\ref{eq:DeBoer_2}) expresses contractile properties of the
myocardium in accordance with Starling's law that takes into
account mechanical effect of the circulation on the BP
(see~\cite{DeBoer:87} and e.g.~\cite{Milnor:89}). The last
equation (\ref{eq:DeBoer_3}) includes explicitly two mechanisms of
the cardiovascular control defined by their respective gain ($G$)
and delay ($\tau$): (i) fast vagal control of the heart rate
$G_vS'_{i-\tau_v}$, (ii) slower $\beta$-sympathetic control of the
heart rate $G_{\beta}F(S',\tau_{\beta})$.
%, and (iii) and slower
%$\alpha$-sympathetic control of the peripheral resistance (or the
%arterial decay time, since arterial compliance is assumed to be
%constant) $G_{\alpha}F(S',\tau_{\alpha})$.
Here $F(S',\tau)$ is a linear weighted sum of the form
\begin{eqnarray*}
 &&F(S',\tau)=\sum_{k=-M}^M a_kS'_{i-\tau+k}=(S'_{i-\tau-2}+
 2S'_{i-\tau-1})\\
 &&+3S'_{i-\tau}+2S'_{i-\tau+1}+S'_{i-\tau+2})/9
 \end{eqnarray*}
Further we assume for simplicity that the pressure oscillations do
not deviate far away from the working point $S_0$ in
(\ref{eq:baroreflex}), i.e. $S' \approx S$.

\begin{figure}
\includegraphics[width=8.cm,height=6cm]{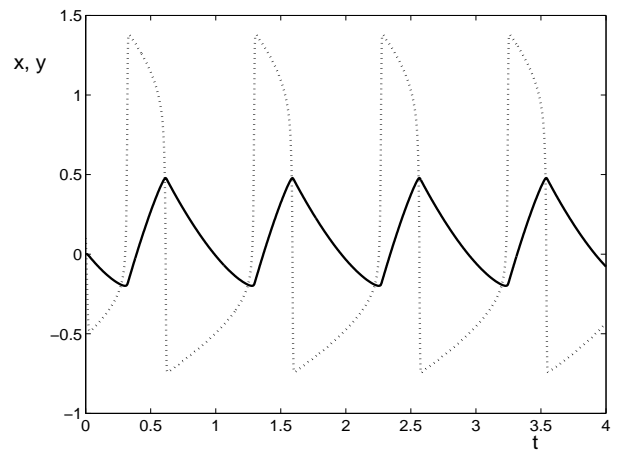}
\caption{\label{fig:fhn} Time evolution of the dynamical variables
$x$ (solid line) and $y$ (dashed line) of the FHN system with the
following parameters: $\epsilon=0.01$, $\beta=-0.05$, $C=-0.125$,
$\alpha=0.5$, $\gamma=1$, $\delta=-1$.}
\end{figure}

To establish the connection between DeBoer (\ref{eq:DeBoer_1}) -
(\ref{eq:DeBoer_3}) model and the model (\ref{eq:2vdp_1}),
(\ref{eq:2vdp_2}) introduced in this paper we note that the
equations DeBoer model is a piece-wise approximation of the actual
BP signal. In particular, it describes the BP signal as an
exponential decay during 2/3 part of the $RR$ interval and linear
increase during 1/3 of the $I_n$ as show in the Fig.
\ref{fig:DeBoer_1} (b).
%---------------------------------------
%------------------------------------------
We note also that the model of the cardiac oscillations
(\ref{eq:2vdp_2}) resembles a model of the FitzHugh-Nagumo (FHN)
system
\begin{eqnarray}\label{eq:FHN}
 \left\{%
\begin{array}{ll}
    \dot x & =  \epsilon (y-\beta x), \\
    \dot y & = \alpha y + \gamma y^2 + \delta y^3 - x + C, \\
\end{array}%
\right.
\end{eqnarray}
where we have neglected for a moment the cardiorespiratory
interaction. The approximation of the BP signal by the output of
the FHN system is also shown in the Fig. \ref{fig:DeBoer_1} (b).
It can be seen already from a comparison between two
approximations that there is a close connections between DeBoer
model and model of coupled oscillators considered in this paper.
This can be further illustrated by noticing that for small
$\epsilon$ the limit cycle in the FHN system consists of fast
motion with practically constant values of $y$, when $x$ jumps
between negative and positive values, and slow motion, when $x$
changes very little (see Fig.~\ref{fig:fhn}). Assuming the
constant value of $x$ at the top $|a_+|$ and at the bottom
$-|a_-|$ of the dashed curve that correspond to the slow motion
along the limit cycle we can integrate the first equation in
(\ref{eq:FHN}) to obtain
%---------------------------------------
%alpha  = 0.5; c      = -0.125; beta   = -.1; gamma  = -1; eta    =
%1; eps    = 2; sc=0.005; a = sc*[0 beta eps];
%a = sc*[0 beta eps]; b = [c alpha eta  gamma  -1];
%out1 = [a(1)+a(2)*y(1)+a(3)*y(2)
%        b(1)+b(2)*y(2)+b(3)*y(2)*y(2)+b(4)*y(2)^3 + b(5)*y(1)];
%------------------------------------------
\begin{eqnarray*}
x_0(t) = \left\{%
\begin{array}{ll}
    (S_{n-1}-|a_-|)e^{-\beta t}+|a_-|, & \hbox{for $0<t<\frac{2}{3}I_{n}$;} \\
    (D_n+|a_+|)e^{-\beta t}-|a_+|, & \hbox{for $\frac{2}{3}I_{n}<t<I_{n}$.} \\
\end{array}%
\right.
\end{eqnarray*}
This solution closely resembles eqs. (\ref{eq:DeBoer_1}) and
(\ref{eq:DeBoer_2}) of the DeBoer model

It can be seen even from this simplified discussion that the
parameters of the model (\ref{eq:2vdp_1}), (\ref{eq:2vdp_2}) found
in the present paper can be related directly to the physiological
parameters of the autonomous control of circulation. Furthermore,
this discussion suggests that it should be possible at least in
principle to bridge inverse and forward modelling and to infer
parameters of the autonomous nervous control of the cardiovascular
system directly from the time-series data.

We emphasize, however, that the obtained results is only the first
step in this direction. In particular, the DeBoer model itself has
to be modified in various ways, including more realistic
functional form of the feedback terms and specifically taking into
account the fact that the baroreflex control is a closed
loop~\cite{Sato:99,Malpas:01}. In fact it was
shown~\cite{Heldt:02} that a multi-compartment closed-loop model
of the cardiovascular responses can simulate well the
experimentally observed variations in the time-series. On the
other hand, this comparison suggests that the inference scheme
used in this paper has to be modified in a various ways to
facilitate convergence and guarantee deeper physiological meaning
of the model parameters as will be discussed in more details
elsewhere. It is also important to emphasize that dynamical
inference of more sophisticated multi-dimensional models of the
type~\cite{Heldt:02} can be addressed only in the frame of full
Bayesian inference of hidden dynamical variables.

\section{Conclusion}
 \label{s:conclusions}

In the present paper we have introduced a technique for nonlinear
dynamical inference of cardiovascular interactions from blood
pressure time-series data. The method is applied to the
simultaneous estimation of the dynamical couplings and noise
strengths in a model of the nonlinear cardio-respiratory
interaction. We have identified a simple nonlinear stochastic
dynamical model of the cardiorespiratory interaction that
describes, in framework of inverse modelling, the time-series data
in a particular frequency band. The method was validated using
surrogate data obtained by numerically integrating the inferred
model itself. We showed that main source of errors in the method
is the decomposition of the blood pressure signal into two
oscillatory components.  We illustrate in the discussion that the
dynamical model of the cardiorespiratory interaction identified in
the present research can be related to the well-know beat-to-beat
model of the cardiovascular control introduced by DeBoer and
co-workers~\cite{DeBoer:85}. The method introduced in this paper
can be used to infer parameters of stochastic nonlinear dynamical
models from observed phenomena across many scientific disciplines.

%\tableofcontents

%\begin{thebibliography}{10}

%\bibliographystyle{prsty}  \bibliography{joint,cardio,bayes_1,optics}

\end{document}